# Growth of Strongly Biaxially Aligned MgB$_2$ Thin Films on Sapphire by Post-annealing of Amorphous Precursors

A. Berenov[1], Z. Lockman[1], X. Qi[1], Y. Bugoslavsky[2][i], L.F. Cohen[2], M.-H. Jo[3], N.A. Stelmashenko[3], V.N. Tsaneva[3], M. Kambara[3], N. Hari Babu[3], D.A. Cardwell[3] M. G. Blamire[3], and J. L. MacManus-Driscoll[1]

[1] Dept. of Materials, Imperial College, Prince Consort Rd, London, SW7 2BP, UK

[2] Dept of Physics, Imperial College, Prince Consort Rd, London, SW7 2BZ, UK

[3] IRC in Superconductivity, University of Cambridge, Madingley Rd., Cambridge, CB3 OHE, UK

**Abstract**

MgB$_2$ thin films were cold-grown on sapphire substrates by pulsed laser deposition (PLD), followed by post-annealing in mixed, reducing gas, Mg-rich, Zr gettered, environments (pO$_2$~10$^{-24}$atm.) at 750°C and 950°C. The films had T$_c$s in the range 29K to 34K, J$_c$s (20K, H=0) in the range 3x10$^4$A.cm$^{-2}$ to 3x10$^5$A.cm$^{-2}$, and irreversibility fields H$^*$ at 20K of 4 T to 6.2 T. An inverse correlation was found between T$_c$ and H$^*$. The films had grain sizes of ~0.1-1μm and a strong biaxial alignment was observed in the 950°C annealed film. (111) oriented MgO was also observed. Mg coating of films during crystallisation appeared to improve film T$_c$.

---

[i] On leave of absence from General Physics Institute, Moscow, Russia

The discovery of superconductivity at 39K in $MgB_2$ (1) has created excitement about the possibility of using this material in magnet (e.g. MRI) and electronic applications (e.g. microwave filters, and SQUID's). The indications of strongly coupled grains in randomly aligned and even impure samples (2, 3) makes $MgB_2$, for low temperature applications, more attractive than high $T_c$ materials which suffer from weakly linked grain boundaries in polycrystalline form. So far, bulk samples have not shown values of the irreversibility field, H*(T), which are high enough to compete with NbTi and $Nb_3Sn$, although recent proton irradiation results have shown that H* can be increased significantly with only moderate damage (4). This is promising for conductor applications where there is the possibility to improve the $J_c$ versus field behaviour through light doping. Recent field anisotropy studies of thin films support the irradiation work. In films with oxygen contamination, $H_{c2}$ (// to a,b) values can be enhanced to ~39T (at 0K) compared to the bulk value of ~ 17T (5).

So far, virtually all thin films have been fabricated by post-annealing of precursors. However, there have been a few reports of $MgB_2$ crystallised during growth, at temperatures as low as 450°C (6). However, $T_c$ is reduced to ~25K. Post-annealed films grown from sintered $MgB_2$ targets generally have $T_c$'s of ~30K (7, 8, 9). Only films grown by e-beam evaporation of boron then post annealed in the presence of Mg vapour, i.e. where oxygen is minimised, give bulk $T_c$ values (10).

Samples have not been grown successfully at high temperatures because Mg has a high volatility, leading to loss of stoichiometry and formation of higher order borides (e.g. $MgB_4$). Thermodynamic calculations indicate that for deposition temperatures of 1000°C, a Mg partial pressure of 340mTorr is required (11). Hence, only those deposition techniques can be used where a high Mg flux from the deposition source can be delivered.

In this work, $MgB_2$ films were deposited on unheated sapphire substrates by PLD from a stoichiometric target prepared from commercial $MgB_2$ powder by pressing and sintering. Approximately 500nm thick films were deposited at a repetition rate of 10Hz at a pressure of 3 Pa in a 4%$H_2$ in Ar gas mixture. The as-deposited films were conducting, but showed activated electrical transport with resistance which increased monotonically as the temperature was reduced to 4.2K.

Films were post-annealed in accordance with the protocols shown in Table 1. Oxygen partial pressure ($pO_2$) was kept low by using reducing gas mixtures, and by using Zr foil-gettering of oxygen. Both Mg powder and foil were also used in the vicinity of the films. The heating and cooling rates used were ~20°C/min, and the dwell time at the peak temperature was 15min. Gas flow rates were ~0.2l/min. The oxygen partial pressure ($pO_2$) was measured using a zirconia sensor at the outflow of the furnace. It is worth noting that the values achieved were still well above the MgO decomposition potential which is $<10^{-50}$ atm. $O_2$ at 750°C.

Depending on the arrangement of the Zr and Mg foils, different amounts of Mg were vaporised onto the surfaces of the films during the anneal. We note that with $H_2$ gas present in the annealing atmosphere, the $pO_2$ depends on the amount of $H_2O$ also present, due to the dynamic equilibrium between $H_2$ and $O_2$ to form $H_2O$.

Magnetisation measurements were performed using a vibrating sample magnetometer (OI-3001, Oxford Instruments). Film surface morphologies were studied by scanning electron microscopy (SEM), and texture by x-ray diffraction (XRD).

Critical temperatures were obtained from magnetic moment versus temperature measurement, m(T). This was done while warming each sample at a field of 1 mT applied perpendicular to the film, the sample having previously been cooled in zero field. The onset $T_c$s are listed in Table 2. The critical current densities, $J_c$s, were evaluated from the width of the magnetisation loops (m(H)) using the Bean model (12). At 10 K, the $J_c$s at zero applied field approached 1 $MA/cm^2$; at 20 K the current densities of the three 'best' films are in excess of 100 $kA/cm^2$ at zero field, and stay above 20 $kA/cm^2$ up to 2 Tesla (see Fig.1).

We performed a length-scale analysis (13) on the highest $J_c$ film, and so estimated the size of the continuous screening current loops within the film. The results shown in Fig. 2 (film 6; 10 K) demonstrate that the current lengthscale is comparable to the film dimensions, and decreases only insignificantly with the field. This confirms that the film is largely free of weak links.

A parameter that is less sensitive to geometric uncertainties is the irreversibility field, $H^*$, which we define as the field where $J_c$ falls below 1 kA/cm$^2$. The values of $H^*$ at 10 and 20 K are also given in Table 2. We plot in Fig. 3 the $H^*$ data on our films at 20K together with the data of Eom et al [7] and for proton-irradiated bulk MgB$_2$ fragments. Among the films, there is a consistent inverse-correlation between $T_c$ and $H^*$. As shown in Fig 3, the value of $H^*$ in film 6 is comparable to that of the optimally-irradiated sample. The depression of $T_c$ in the film is however much larger, thus suggesting that the pinning mechanism, or the defects responsible for pinning are completely different in the two cases.

Despite the very reducing nature of the anneals, we did not produce a 39K $T_c$ film, even at the highest annealing temperature of 950°C where sufficient structural ordering should occur. Indeed, the highest annealing temperature did not produce the highest $T_c$ film. The highest $T_c$ measured was 34K (for film 3). The film was neither crystallised under a more reducing condition (e.g. compared to film 2) nor at a higher temperature (e.g. compared to film 8) than the other films. However, the film was almost completely covered in bright Mg which evaporated from the Mg source material in its immediate vicinity. Two possible reasons for the higher $T_c$ of film 3 are: the surface Mg layer is gettering oxygen from the film or annealing atmosphere, leading to a lower oxygen content, or excess Mg in the surface layer compensates for the lowering of the overall Mg stoichiometry due to MgO formation. Hence, if lower $T_c$s arise from a slightly Mg-deficient MgB$_2$ phase, the $T_c$ should be higher in the presence of a Mg excess.

Fig. 4 shows an x-ray diffractogram for film 8. Sharp MgB$_2$ (001) peaks are observed, as well as Al$_2$O$_3$ (001) peaks (the highest intensity Al$_2$O$_3$ (006) substrate peak at 2θ ~41.7°was removed from the plot). MgO (111) and (222) peaks are also observed, their strong intensities relative to the MgO (220) peak at 62.3° indicating (111) orientation of the phase. Barely detectable MgB$_2$ peaks were seen for the 750°C films indicative of poorer crystallinity and/or random orientation at the lower anneal temperature. Because of the MgO orientation, it is not possible to compare the level of MgO found with that in the films of Eom et al. which were annealed in a relatively strong oxidising atmosphere (7). However, based on their more highly oxidising annealing atmosphere, we expect much less MgO. Since the films were grown from a stoichiometric target and since there is some phase fraction of MgO, there should be a small quantity of other boride phases present. From thermodynamic calculations, in the presence of only $10^{-23}$ atm. $CO_2$ and for our chosen annealing conditions, in addition to MgO and $MgB_4$, $B_4C$ is also a stable phase in equilibrium with $MgB_2$ (14).

For a few films, we also undertook $\phi$ scans of the $MgB_2$ (102) peaks at $2\theta=63.337°$, and $\Phi=33.4°$ ((102) is the only peak which does not arise at a Bragg angle very close to one of the $Al_2O_3$ peaks). The 750°C annealed films showed no detectable in-plane alignment.

Fig. 5 shows $\phi$ scans for the 950°C annealed, film 8. Fig. 5a was undertaken at $2\theta=61.3°$, $\Phi=33.4°$, for the $Al_2O_3$ (108) planes. A 3 fold symmetry is observed, as expected. Fig. 5b was undertaken at $2\theta=63.33°$, $\Phi=33.4°$. Six main peaks are observed as well as three minor peaks. In order to determine the origin of the two different sets of peaks, $\theta$-$2\theta$ scans were undertaken on one peak in each set. For the main set, a sharp, symmetrical peak was found at a $2\theta$ value of 63.3°, indicative of $MgB_2$, whereas for the minor set, a low intensity shoulder was observed, indicative of a tail from one of the $Al_2O_3$ (108) peaks. Hence, the six fold symmetry of the $MgB_2$ peaks indicates that a single, sharp, in-plane orientation occurs for the $MgB_2$ grains on sapphire. The angular difference between the $Al_2O_3$ <100> axis and $MgB_2$ <100> axis in the overlying grains is 30°.

The grain sizes in the films varied from of ~0.1-1μm with no systematic dependence of grain size or morphology on annealing atmosphere. Film 8 annealed at 950°C had a grain size of ~0.1μm, i.e. not larger than for the 750°C films, as might be expected, indicative of a grain growth blocking effect, probably by MgO particles. The grain sizes are larger by around an order to magnitude compared to those of Eom et al. (7). Also, the grain boundaries are more pronounced in our films compared to similarly processed films by Zhai et al. (10).

In summary, five PLD, cold-grown, amorphous $MgB_2$ films were crystallised in a Mg-rich environment, reducing atmosphere ($pO_2$'s ~ 1.8-3x$10^{-24}$atm), at temperatures of 750°C and 950°C. $T_c$'s of between 29K and 34K were achieved with $J_c$'s (20K, H=0) in the range 3x$10^4$A.cm$^{-2}$ to 3x$10^5$A.cm$^{-2}$, and H$^*$ at 20K between 3.9T and 6.2 T. An inverse correlation was found between $T_c$ and H$^*$. The pinning defects in the films are clearly different to those in irradiated bulk fragments. Oxygen presence in both the deposition and annealing stages plays an important role in film properties, whether this be through alloying, and/or grain size control associated with MgO formation. The film with the highest $T_c$ (~34K) was almost completely covered with surface Mg which had evaporated and condensed out of the nearby Mg source. The 950°C annealed film showed strong biaxial alignment of grains which is encouraging for growth of highly aligned films at much lower temperatures by in-situ growth.

For $MgB_2$ film fabrication for device applications, use of single Mg and B targets should be beneficial over composite $MgB_2$ targets since this will allow reduced oxygen incorporation in the films. Also, during the crystallisation stage, highly gettered annealing environments will be required so as to reduce the rate of MgO formation.

**Acknowledgments**

The authors would like to thank the EPSRC for funding this work. Also, helpful discussions with Prof. D. Larbalestier are gratefully acknowledged.

| Sample number | T (°C) | Orientation of sapphire substrate | Atmosphere | $pO_2$ in annealing gas (atm.) |
|---|---|---|---|---|
| 1 | 750 | r-plane | 95%Ar,5%$H_2$ | $2.1 \times 10^{-24}$ |
| 2 | 750 | c-plane | 90%$N_2$,10%Ar | $1.8 \times 10^{-24}$ |
| 3 | 750 | c-plane | 95%Ar,5% $H_2$ | $2.1 \times 10^{-24}$ |
| 6 | 750 | c-plane | 98%Ar,2% $H_2$ | $3.0 \times 10^{-24}$ |
| 8 | 950 | c-plane | 98%Ar,2%$H_2$ | $3.0 \times 10^{-24}$ |

Table 1: Film post-annealing conditions

| Sample number | $T_c$ (K) | $J_c$ (T=20K, H=0) A/cm$^2$ | $H^*$ (T) (T=10 K) | $H^*$ (T) (T=20 K) | Comments |
|---|---|---|---|---|---|
| 1 | 31.8 | $2.7 \ 10^5$ | 5 | 3.9 | Low, even Mg coverage |
| 2 | 30.5 | $1.7 \ 10^5$ | >8 | 4.9 | Low, even Mg coverage |
| 3 | 34.0 | $3.0 \ 10^4$ | 5.9 | 4.0 | Almost complete Mg coverage |
| 6 | 29.2 | $2.3 \ 10^5$ | >8 | 6.2 | Low, patchy Mg coverage of film surface |
| 8 | 31.9 | $9 \ 10^4$ | 7.9 | 4.4 | Sample wrapped in Zr foil and then Mg foil. No Mg coverage of film surface |

Table 2: Superconducting properties of post-annealed films

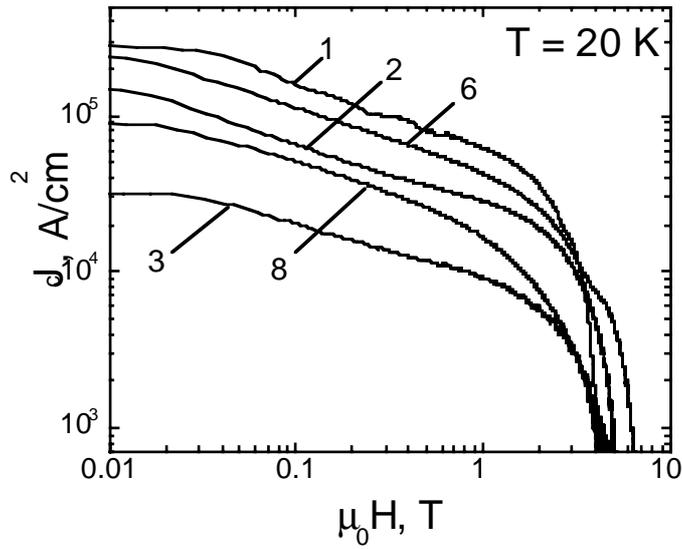

Figure 1. Critical current density ($J_c$) as a function of the applied magnetic field ($\mu_o H$) at T=20 K. The data were inferred from the magnetisation hysteresis loops using the Bean model, and assuming the sample to be fully-connected.

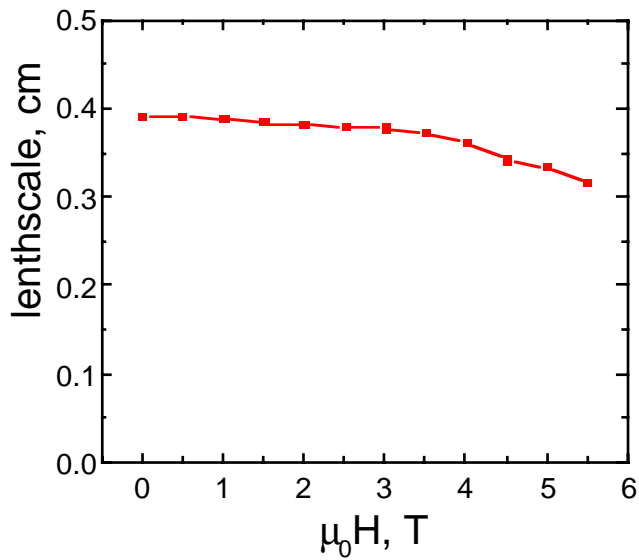

Figure 2. The characteristic current-flow length-scale in film 6, at T=10 K, which is comparable with the sample half-width of 4x6mm, indicating that the sample is fully-connected

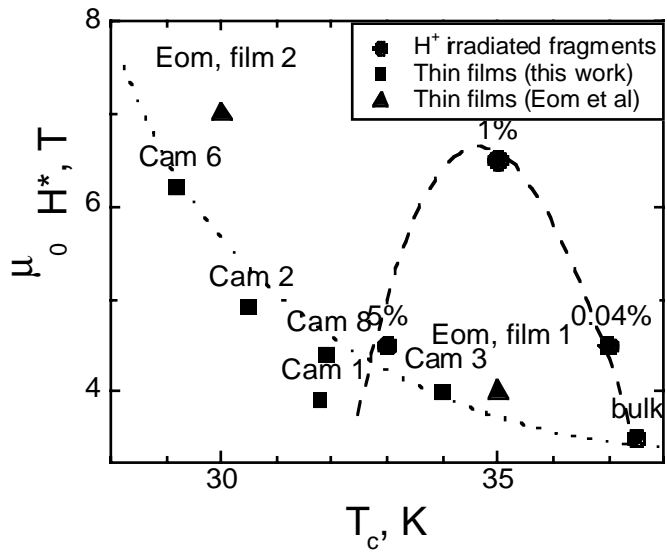

Figure 3. Correlation between $T_c$ and irreversibility field $H^*$ at T=20 K for various samples.

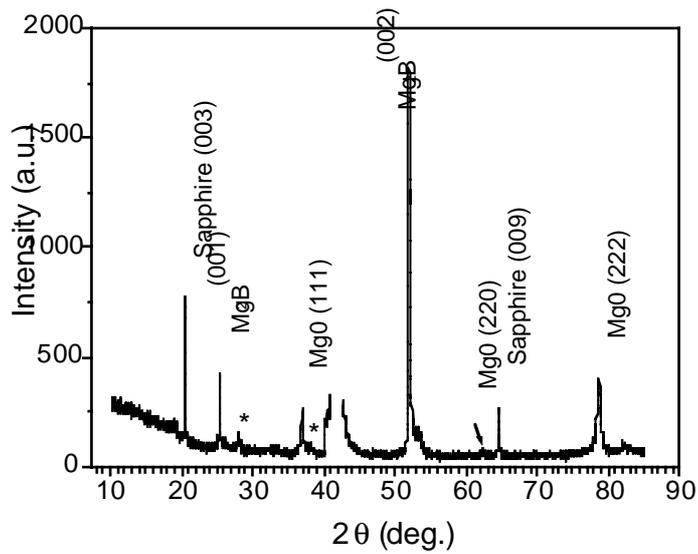

Figure 4: $\theta$-$2\theta$ x-ray scan for film 8 showing sharp (00l) peaks of $MgB_2$. Asterisks are unknown peaks.

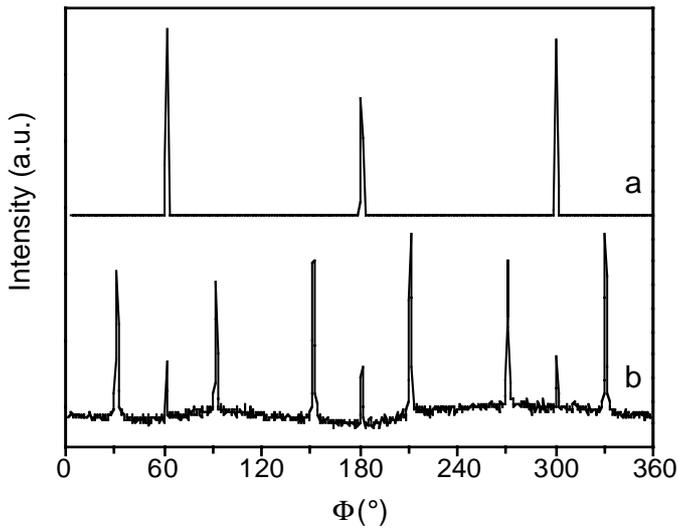

Figure 5: X-ray phi scans for (a) (108) $Al_2O_3$ planes and (b) (102) $MgB_2$ planes (major peaks at 60° intervals from initial peak at 30°) + (108) $Al_2O_3$ planes (minor peaks at 120° intervals from initial peak at 60°). The six fold symmetry indicates strong in-plane texture of the $MgB_2$ grains.